\documentclass{article}

\usepackage[preprint]{neurips_2021}

\usepackage[utf8]{inputenc} 
\usepackage[T1]{fontenc}    
\usepackage{hyperref}       
\usepackage{url}            
\usepackage{booktabs}       
\usepackage{amsfonts}       
\usepackage{amsmath}
\usepackage{mathtools}
\usepackage{nicefrac}       
\usepackage{microtype}      
\usepackage{xcolor}         
\usepackage{multirow}
\usepackage{graphicx}
\usepackage{bm}
\usepackage{rotating}

\title{COmic: Convolutional Kernel Networks for Interpretable End-to-End Learning on (Multi-)Omics Data}

\author{%
  \hspace{1mm}Jonas C.~Ditz \\
	Methods in Medical Informatics\\
	Department of Computer Science\\
	University of T\"{u}bingen\\
	T\"{u}bingen, Germany \\
	\texttt{jonas.ditz@uni-tuebingen.de} \\
	\And
	\hspace{1mm}Bernhard Reuter \\
	Methods in Medical Informatics\\
	Department of Computer Science\\
	University of T\"{u}bingen\\
	T\"{u}bingen, Germany \\
	\texttt{bernhard.reuter@uni-tuebingen.de} \\
	\And
	\hspace{1mm}Nico Pfeifer \\
	Methods in Medical Informatics\\
	Department of Computer Science\\
	University of T\"{u}bingen\\
	T\"{u}bingen, Germany \\
	\texttt{nico.pfeifer@uni-tuebingen.de} \\
}

\begin{document}

\maketitle

\begin{abstract}
  \textbf{Motivation:} The size of available omics datasets is steadily increasing with technological advancement in recent years. While this increase in sample size can be used to improve the performance of relevant prediction tasks in healthcare, models that are optimized for large datasets usually operate as black boxes. In high stakes scenarios, like healthcare, using a black-box model poses safety and security issues. Without an explanation about molecular factors and phenotypes that affected the prediction, healthcare providers are left with no choice but to blindly trust the models. We propose a new type of artificial neural network, named Convolutional Omics Kernel Network (COmic). By combining convolutional kernel networks with pathway-induced kernels, our method enables robust and interpretable end-to-end learning on omics datasets ranging in size from a few hundred to several hundreds of thousands of samples. Furthermore, COmic can be easily adapted to utilize multi-omics data.\\
  \textbf{Results:} We evaluated the performance capabilities of COmic on six different breast cancer cohorts. Additionally, we trained COmic models on multi-omics data using the METABRIC cohort. Our models performed either better or similar to competitors on both tasks. We show how the use of pathway-induced Laplacian kernels opens the black-box nature of neural networks and results in intrinsically interpretable models that eliminate the need for \textit{post-hoc} explanation models.\\
  \textbf{Availability:} Datasets, labels, and pathway-induced graph Laplacians used for the single-omics tasks can be downloaded \href{https://ibm.ent.box.com/s/ac2ilhyn7xjj27r0xiwtom4crccuobst/folder/48027287036}{here}. While datasets and graph Laplacians for the METABRIC cohort can be downloaded from the above mentioned repository, the labels have to be downloaded from \href{https://www.cbioportal.org/study/clinicalData?id=brca_metabric.}{cBioPortal}. COmic source code as well as all scripts necessary to reproduce the experiments and analysis are publicly available at \href{https://github.com/jditz/comics}{https://github.com/jditz/comics}.
\end{abstract}

\section{Introduction}
In recent years, artificial neural networks (ANNs) show promising performance when employed to learn correlations between data points and outcome variables. They combine feature extraction and prediction training in a single end-to-end learning scheme lowering the necessary amount of labour put into feature engineering and can be used on very large datasets with relative ease. With the advent of big data and high-throughput data generation techniques in computational biology and healthcare, resulting in an increased number of data points available for the training of prediction models, the use of ANNs in these fields has vastly increased. In computational biology, ANNs showed promising performance capabilities when applied to prediction tasks in regulatory genomics {\citep{alipanahi2015predicting,zhou2015predicting,kelley2016basset}} and in biological image analysis {\citep{ronneberger2015u,parnamaa2017accurate,ferrari2017bacterial}}. Furthermore, several authors showed the potential of ANNs in healthcare scenarios such as diagnosis {\citep{arieno2019review,de2018clinically}}, drug discovery {\citep{alvarez2019using,fleming2018artificial}}, epidemiology {\citep{hay2013big}}, personalized medicine {\citep{miotto2016deep}}, and operational efficiency {\citep{nelson2019predicting}}. However, utilizing ANNs for prediction tasks usually comes with two shortcomings: First, a large amount of data is needed to robustly train a deep neural network and, second, neural network models operate as black-boxes. While the first problem can be tackled by shrinking the complexity of the neural network, which increases the stability of the model but often leads to a decrease in performance, the second shortcoming is most often addressed using \textit{post-hoc} interpretation models. This technique involves solving a secondary task that utilizes a pre-trained prediction model such that the computed solution provides a humanly understandable interpretation for the results computed by the prediction model. Commonly used methods include Shapley additive explanation (SHAP, \cite{lundberg2017unified}), counterfactual explanation using generative models {\citep{stepin2021survey}}, and saliency methods like Layer-wise Relevance Propagation (LRP, \cite{montavon2019layer}), Deep Taylor Decomposition (DTD, \cite{montavon2017explaining}), GuidedBP {\citep{springenberg2014striving}}, or DeepLIFT {\citep{shrikumar2017learning}}. Using \textit{post-hoc} interpretation methods can provide additional information to improve understanding and advance scientific knowledge in low-risk scenarios but they have several properties that render their use in high-risk scenarios potentially problematic. Most \textit{post-hoc} interpretation methods are unfaithful to the computations of the original model {\citep{rudin2019stop}}. Furthermore, many saliency methods ignore information provided by deeper layers of ANNs {\citep{sixt2020explanations}}. Recent work showed that \textit{post-hoc} methods are limited in adversarial contexts {\citep{bordt2022post}} and can be exploited to provide seemingly plausible but misleading explanations {\citep{lipton2018mythos}}. In healthcare, decisions that are made based on wrong or misleading explanations have the potential to cause harm to patients.

Kernel methods can provide both robustness on small datasets and interpretation capabilities within the domain of the data. These methods utilize the kernel trick to solve a prediction task by implicitly projecting data into the reproducing kernel Hilbert space (RKHS) of a kernel function and solve the classification or regression problem within the RKHS. While the use of a kernel functions does not always guarantee interpretation capabilities, there are several kernel functions for biological data that result in interpretable models, e.g., the oligo kernel for sequences {\citep{meinicke2004oligo}} or the pathway-induced kernel for omics data {\citep{manica2019pimkl}}. Combining kernel functions with ANNs is a promising direction to increase the robustness of ANN models on small datasets and several efforts in that direction have been published in recent years {\citep{cho2009kernel,bo2011object,mairal2014convolutional,mairal2016end}}. Chen and colleagues showed the feasibility of kernel networks for biological sequences by using a relaxation of the mismatch kernel {\citep{eskin2003mismatch}} to build convolutional and recurrent neural network architectures {\citep{chen2019biological,chen2019recurrent}}. Furthermore, they showed how to use convolutional kernel neural networks on graph-structured data like protein structures {\citep{chen2020convolutional}}. While these models showed promising results and increased robustness, the choice of the kernel function resulted in models that are not intrinsically interpretable. However, we recently showed that a carefully chosen kernel function results in intrinsically interpretable kernel networks for biological sequence data {\citep{ditz2021convolutional}}. With this work we introduce Convolutional Omics Kernel Networks (COmic), a neural network architecure that allows for intrinsically interpretable end-to-end learning on (multi-)omics data. This is achieved by using a kernel function based on graph Laplacians of biological networks to project input samples into a subspace of the corresponding reproducing kernel Hilbert space (RKHS) with a variant of the Nystr\"{o}m method. Using max-pooling combined with strictly linear layers for classification results in COmic models that provide global interpretation, while attention layers can be used to create COmic models that provide local interpretation. In this manuscript, we use the definition most commonly found in the interpretable ML literature for global and local interpretation {\citep{molnar2020interpretable}}. In simple words, \textit{global interpretation} can be used to answer the question \textit{"How does the trained model make predictions?"}, while \textit{local interpretation} can be used to answer the question \textit{"Why did the model make a certain prediction for a specific input?"}. 

We show the performance and interpretation capabilities of COmic models on six different breast cancer microarray cohorts. These cohorts contain microarray gene expression data from patients with breast cancer and were stratified on the occurrence of a relapse within 5 years. We compare our proposed method to 15 previously published approaches including several methods based on support vector machines (SVMs) like network-based SVMs {\citep{zhu2009network}}, recursive feature elimination SVMs {\citep{guyon2002gene}}, and graph diffusion kernels for SVMs {\citep{rapaport2007classification,gao2009graph}} as well as classification by average pathway expression {\citep{guo2005towards}}, classification by significant hub genes {\citep{taylor2009dynamic}}, classification by pathway activity {\citep{lee2008inferring}}, and pathway-induced multiple kernel learning (PIMKL, \cite{manica2019pimkl}). We show how the projection into a subspace of the RKHS of pathway-induced kernels in combination with linear and attention layers leads to global and local interpretations, respectively. Furthermore, we use the METABRIC cohort {\citep{curtis2012genomic}} to show how COmic models can be used on multi-omics data. On the METABRIC breast cancer cohort, we predicted disease-free survival using gene expression (mRNA) and copy number alteration (CNA) data.

This work is structured as follows. We first introduce COmic by describing the pathway-induced kernel and define the necessary network architecture to build a COmic model. Afterwards, we show how to achieve a globally interpretable COmic model using strictly linear layers and a locally interpretable COmic model using attention layers. We evaluate COmic models on six breast cancer cohorts and show how COmic models can be utilized for multi-omics data. With this manuscript, we introduce a new kernel network architecture that can be both robustly trained on small-scale (multi-)omics datasets and easily utilized for prediction tasks on (multi-)omics datasets with several hundreds of thousands of data points. Furthermore, our method results in intrinsically interpretable models offering global and local interpretations of prediction results.


\begin{figure*}
    \centering
    \includegraphics[width=0.8\textwidth]{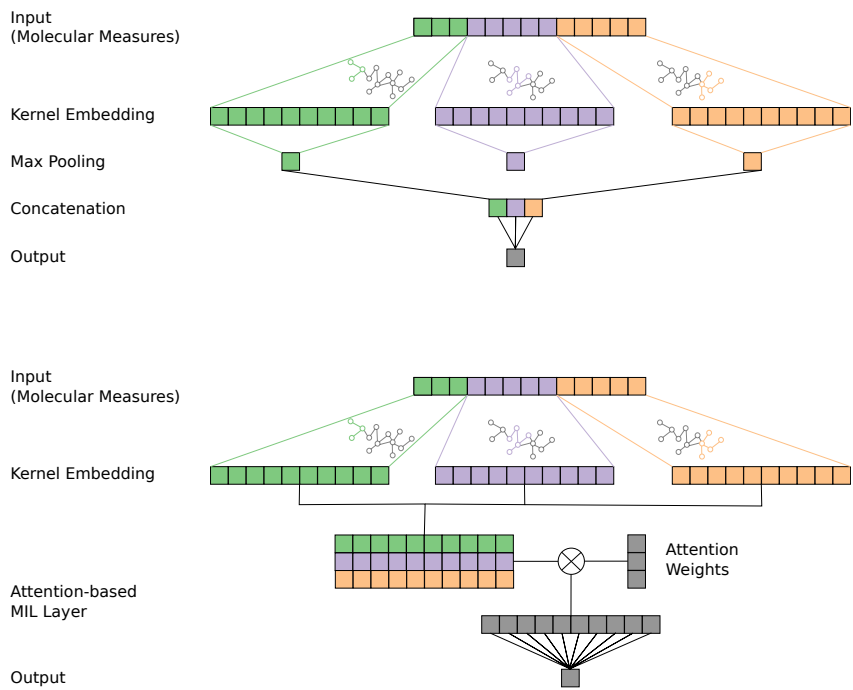}
    \caption{Schematic of the proposed interpretable COmic models. \textbf{Top:} Pooling-based COmic model. The kernel embedding of each involved pathway are reduced to a single dimension by using a one-dimensional max pooling operation. The output of each pooling layer is concatenated and prediction is performed using a strictly linear fully-connected layer. The pooling-based COmic models are globally interpretable similar to PIMKL models by utilizing the weights of the fully-connected layer. \textbf{Bottom:} Attention-based COmic model. The kernel embeddings of each involved pathway are transformed into a bag of instances of a multiple instance learning problem. Attention weights for each instance are calculated using each pathway's kernel embedding and a matrix multiplication between the bag of instances and the attention weights is performed. The output is used for prediction with a strictly linear fully-connected layer. The attention-based COmic models can be locally interpreted by utilizing the attention weights.}
    \label{fig:architecture}
\end{figure*}

\section{Convolutional Omics Kernel Networks}
In the following section, we describe the theoretical background of convolutional kernel networks for prediction tasks on omics-based datasets.

\subsection{Pathway-Induced Kernel Functions}
The foundation of pathway-induced kernel functions are so-called graph Laplacian matrices. To define these matrices we first assume $G = (V, E)$ to be an undirected graph with vertices $V = \lbrace v_1, \ldots, v_n \rbrace$ and edges $E$. Furthermore, $G$ is assumed to be a weighted graph with weight matrix $W \in \mathbb{R}^{n \times n}$, where $w_{ij} = w_{ji} \ge 0$ describes the weight of the edge between vertices $v_i$ and $v_j$. The degree of each vertex $v_i \in V$ is defined as $d_i = \sum_{j=1}^{n} w_{ij}$. The diagonal matrix with the degrees $d_1, \cdots, d_n$ on the diagonal is called the degree matrix $D$. The unnormalized graph Laplacian $L \in \mathbb{R}^{n \times n}$ is defined as {\citep{von2007tutorial}}:
\begin{equation}
    \label{eq:unnormLap}
    L \coloneqq D - W
\end{equation}
Since an unnormalized graph Laplacian has undesirable mathematical properties in case of very broadly distributed degrees within $G$ {\citep{von2007tutorial}}, we use a normalized graph Laplacian instead, which is defined as:
\begin{equation}
    \label{eq:normLap}
    L_{\text{sym}} \coloneqq D^{-\frac{1}{2}} L D^{-\frac{1}{2}}
\end{equation}

Similar to previous manuscripts, see e.g., {\citep{chen2011identifying,manica2019pimkl}}, we use molecular interaction networks (MIN) as graphs underlying the normalized graph Laplacians. Using known interaction networks allows to define a kernel function that computes the similarity of molecular measures (gene expression, DNA methylation, etc.) under the assumed interactions defined by the network. Given two molecular measures $x_i \in \mathbb{R}^n$ and $x_j \in \mathbb{R}^n$, we define the kernel function as
\begin{equation}
    \label{eq:kernel}
    K_{\text{MIN}}(x_i, x_j) = x_i^T L_{\text{MIN}} x_j,
\end{equation}
where $L_{\text{MIN}}$ is the normalized graph Laplacian (as defined in Eq. \ref{eq:normLap}) of a molecular interaction network.

Manica and colleagues proposed to use pathway-specific sub-networks instead of whole interaction networks for computing normalized graph Laplacians {\citep{manica2019pimkl}}. This method allows for a more tailored induction of prior knowledge into a prediction task. The authors call this approach pathway-induced (PI) kernel functions. Here, the similarity between two molecular measures is not computed using a single graph Laplacian but with a set of $p$ graph Laplacians $\mathbf{L} = \lbrace L_{\text{PI}_1}, \ldots, L_{\text{PI}_p} \rbrace$ each defined over a pathway-specific sub-network of the molecular interaction network. Therefore the pathway-induced kernel for two molecular measures $x_i \in \mathbb{R}^n$ and $x_j \in \mathbb{R}^n$ is not a single function but a set of functions defined as
\begin{equation}
    \label{eq:pikernel}
    K_{\text{PI}}(x_i, x_j) = \lbrace K_{\text{PI}_1}(x_{i,\text{PI}_1}, x_{j,\text{PI}_1}), \ldots, K_{\text{PI}_p}(x_{i,\text{PI}_p}, x_{j,\text{PI}_p}) \rbrace
\end{equation}
with
\begin{equation}
    K_{\text{PI}_r}(x_{i,\text{PI}_r}, x_{j,\text{PI}_r}) = x_{i,\text{PI}_r}^T L_{\text{PI}_r} x_{j,\text{PI}_r},
\end{equation}
where $L_{\text{PI}_r} \in \mathbb{R}^{d \times d}$ is the symmetric graph Laplacian of the $r$th pathway-specific sub-network with $d$ nodes and $x_{i,\text{PI}_r} \in \mathbb{R}^d$ and $x_{j,\text{PI}_r} \in \mathbb{R}^d$ are vectors containing only the signal values of the molecular measures that correspond to nodes within the pathway-specific sub-network described by $L_{\text{PI}_r}$. Manica and colleagues used multiple kernel learning (MKL) to combine the set of pathway-induced kernel functions into a single learning framework. In contrast to prior work, we are using a variant of the Nystr\"{o}m method to formulate an explicit parametrization of an orthogonal projection onto a finite-dimensional subspace of a pathway-induced kernel's RKHS. This enables us to define pathway-induced kernel layers that can be incorporated into artificial neural networks. By using a neural network architecture as the basis for COmic, the resulting models can be tailored to specific datasets (single- or multi-omics) as well as the desired form of interpretation. We will show that in the following sections.

\subsection{Convolutional Kernel Layer projects onto a finite-dimensional RKHS-Subspace}
\label{sec:kernelLayer}
Convolutional kernel networks make use of a variant of the Nystr\"{o}m method to project input samples into a finite-dimensional subspace of the RKHS $\mathcal{H}$ of a kernel function. To achieve this, a set of $q$ anchor points $z_1, \ldots, z_q$ is used to define a $q$-dimensional subspace $\mathcal{E}$ of $\mathcal{H}$. The anchor points lie in the input space of the kernel function and the RKHS subspace is defined as
\begin{equation}
    \mathcal{E} = \text{Span}(\phi_{z_1}, \ldots, \phi_{z_q}),
\end{equation}
where $\phi(z_i)$ denotes the image of the $i$th anchor point under the kernel function. The orthogonal projection of input points onto $\mathcal{E}$ admits an explicit parametrization that utilizes the kernel trick to avoid explicitly calculating the images $\phi(z_i)$ {\citep{mairal2016end,williams2001using,zhang2008improved}}. For an input $x$, i.e., a molecular measure in case of omics data, the explicit parametrization $\psi(x) \in \mathbb{R}^q$ is defined as
\begin{equation}
    \psi(x) = K_{ZZ}^{-\frac{1}{2}} K_Z(x),
\end{equation}
where $K_{ZZ} = (K(z_i, z_j))_{i = 1, \ldots, q; j = 1, \ldots, q}$ is the gram matrix formed by the anchor points, $K_{ZZ}^{-\frac{1}{2}}$ denotes the (pseudo-)inverse square root of the Gram matrix, and $K_Z(x) = \left[ K(x,z_1), \ldots, K(x,z_p) \right]^T$. As shown in Figure \ref{fig:architecture}, each pathway-induced kernel function has to be modelled with a separate orthogonal projection. This means that a COmic model utilizing $p$ pathway-induced kernel functions maps each input onto $p$ representations $\psi_{\text{PI}_1}, \ldots, \psi_{\text{PI}_p} \in \mathbb{R}^q$. These representations are then used to solve the prediction task for the input. In the next section, we show two different approaches to combine the representations leading to globally or locally interpretable models, respectively.

The anchor points can be initialized using $k$-means on all input samples with the number of clusters set to the number of anchor points. Afterwards, the anchor points are optimized with the end-to-end learning scheme used to train the whole network. For all experiments described in this manuscript, anchor points were initialized using $k$-means++ {\citep{vassilvitskii2006k}}.

\subsection{Globally and Locally Interpretable COmic Models}
\label{sec:interpret}
\textit{Globally interpretable COmic models} are based on multi-kernel learning (MKL). A simple approach to MKL is finding an optimal linear combination of all utilized kernels. This approach learns a weight for each kernel and, therefore, can be used to determine the influence each kernel has on the prediction outcome. Since kernels are directly associated with pathways in PIMKL, Manica and colleagues show that MKL weights can be used to determine the importance of different pathways for a prediction {\citep{manica2019pimkl}}. We can embed a similar weighted sum of pathway-induced kernels into the architecture of COmic models. Each kernel embedding produced by the PI-kernel layer described in section \ref{sec:kernelLayer} is passed into a one-dimensional max pooling layer. This results in a single activation $A_r = \max \left( \psi_{PI_r} \right)$ for each pathway-induced kernel, where $\max \left( \psi_{PI_r} \right)$ denotes the maximum value in vector $\psi_{PI_r}$. This activation is high, if the input is similar to one of the learned anchor points, and low otherwise. By concatenating all activations and passing them into a strictly linear fully-connected layer, the model learns a single weight for each pathway-induced kernel and the prediction is calculated as a weighted sum of all kernels, i.e.,
\begin{equation}
    \label{eq:pool_decision}
    \hat{y} = \sum_{r=1}^{p} w_r A_r,
\end{equation}
where $\hat{y}$ is the overall prediction, $w_r \in \mathbb{R}$ is the weight and $A_r \in \mathbb{R}$ is the activation of the $r$th pathway-induced kernel. We call this architecture \textit{pooling-based COmic model}. In contrast to the MKL approach, the weights can become negative. This enhances the interpretation capabilities of pooling-based COmic models, since we cannot only infer if a pathway is important for the prediction task but also with which class each pathway is associated by looking at the sign of the weight. The top part of Figure \ref{fig:architecture} shows a schematic of a pooling-based COmic model.

\textit{Locally interpretable COmic models} are based on multiple instance learning (MIL). In MIL, each sample is represented as a bag of instances with a single label per bag {\citep{dietterich1997solving,maron1997framework,oquab2014weakly}}. There are two general approaches to solve an MIL problem: the instance-level approach and the embedding-level approach. In the instance-level approach, an instance-level classifier predicts a score for each of the instances in the bag. Afterwards, scores are aggregated by MIL pooling to compute the prediction for the bag. In the embedding-level approach, a low-dimensional embedding of each instance is computed and MIL pooling is used on the embedded instances to create a bag representation. This representation is used by a bag-level classifier to provide the prediction. While it was shown that the embedding-level approach leads to better performances {\citep{wang2018revisiting}}, the instance-level approach leads to interpretable models {\citep{liu2012key}}. Ilse and colleagues proposed an MIL-model based on neural networks that combines the strength of both approaches, called attention-based multiple instance learning {\citep{ilse2018attention}}. Their approach can be utilized for COmic models in the following way. The output of our proposed PI-kernel layer can be viewed as a bag of low-dimensional instances $H = \{\psi_{PI_1}, \ldots, \psi_{PI_p}\}$, where each $\psi_{PI_r} \in \mathbb{R}^q$ is the projection onto a $q$-dimensional subspace of the RKHS of one pathway-induced kernel. Attention-based MIL pooling is then used to compute the bag representation, i.e.,
\begin{equation}
    \label{eq:bagrep}
    \tilde{\psi} = \sum_{r=1}^{p} a_r \psi_{PI_r},
\end{equation}
where
\begin{equation}
    \label{eq:normal}
    a_r = \frac{\exp \{ w^T \tanh ( V \psi_{PI_r}^T)\}}{\sum_{j=1}^{p} \exp \{ w^T \tanh ( V \psi_{PI_j}^T)\}}.
\end{equation}
$w \in \mathbb{R}^{l \times 1}$ and $V \in \mathbb{R}^{l \times m}$ are parameters of the attention layer. As noticed by Ilse and colleagues, the $\tanh(\cdot)$ non-linearity introduces a potential limitation due to the fact that it is roughly linear only for $x \in [-1, 1]$. This limitation can be reduced by using a gating mechanism {\citep{dauphin2017language}}. In this case, the attention weights are calculated as
\begin{equation}
    \label{eq:gated}
    a_r = \frac{\exp \{ w^T (\tanh ( V \psi_{PI_r}^T) \odot \text{sigm} ( U \psi_{PI_r}^T))\}}{\sum_{j=1}^{p} \exp \{ w^T (\tanh ( V \psi_{PI_j}^T) \odot \text{sigm} ( U \psi_{PI_j}^T)\}}.
\end{equation}
Again, $w \in \mathbb{R}^{l \times 1}$, $V \in \mathbb{R}^{l \times m}$, and $U \in \mathbb{R}^{l \times m}$ are parameters of the attention layer. In both cases, the training of all attention layer parameters is part of the end-to-end training routine for the whole network and, hence, does not introduce the need for additional measures. We call this architecture \textit{attention-based COmic model}. Since the attention weights $a_p$ are input specific, they enhance a model with local interpretation capabilities. The bottom part of Figure \ref{fig:architecture} shows a schematic of an attention-based COmic model.

\begin{figure}
    \centering
    \includegraphics[width=\textwidth]{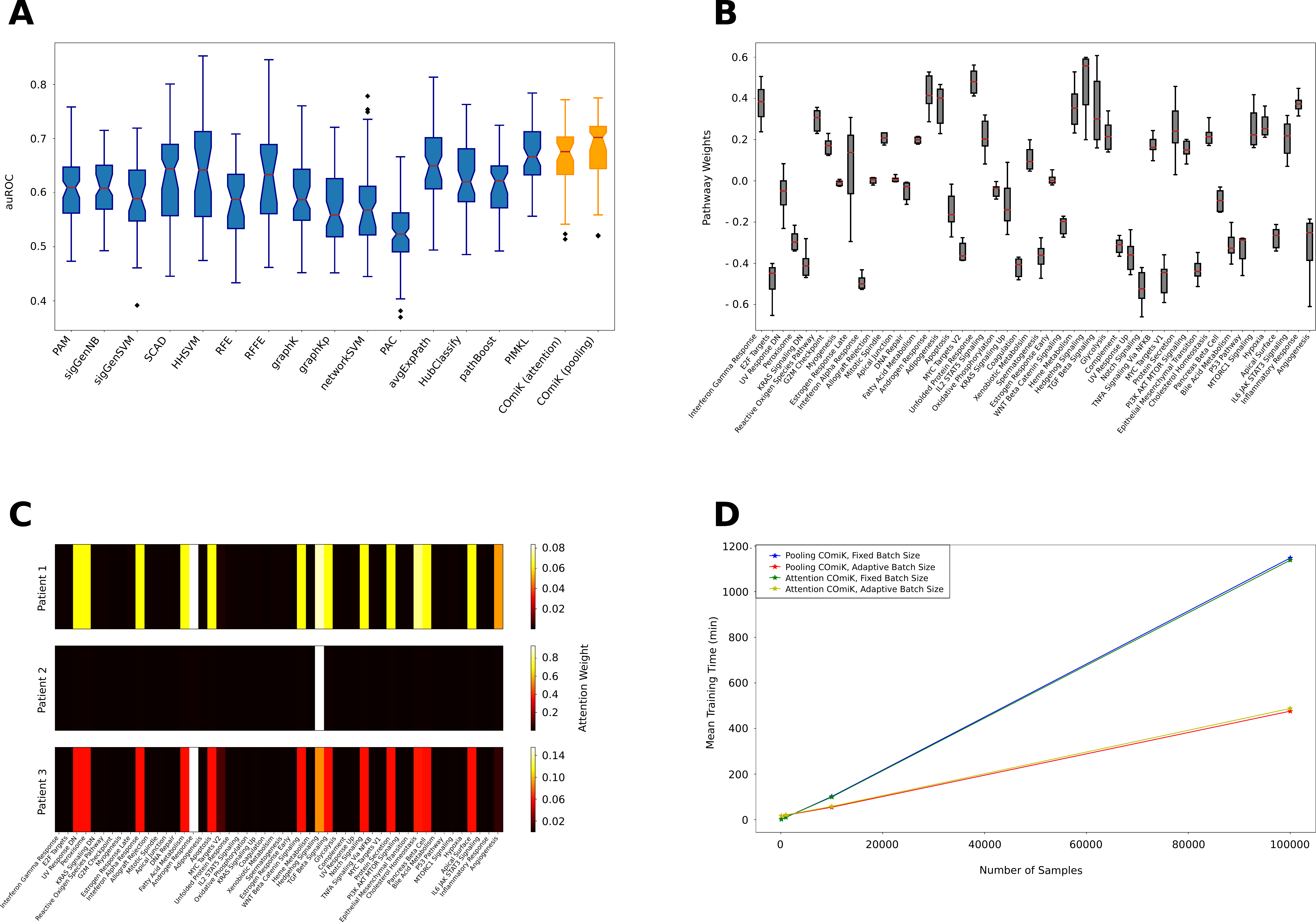}
    \caption{\textbf{A}: Cross-validation performance of COmic models compared to previously published methods. The boxplots show the ten mean auROC validation scores of a 10-times repeated 10-fold cross-validation over each of the six breast cancer cohorts. Performance of COmic models are shown in orange. The center line of each box indicates the median. The height of the boxes represents the inter quartile range (IQR) with the upper and lower whiskers set to 1.5 times the IQR. Outliers are depicted by black diamonds. Notches represent the confidence interval (CI) around the median and were calculated using bootstrapping with 10000 iterations. \textbf{B:} Visualizing the global interpretation capabilities of a pooling-based COmic model. Each box represents one of the 50 pathways and was created using the pathway weights of the models trained on the six publicly available breast cancer cohorts: GSE11121, GSE1456, GSE2034, GSE2990, GSE4922, and GSE7390. The boxplots are defined as in (\textbf{A}) but without notches (CIs not shown). \textbf{C:} Visualizing the local interpretation capabilities of an attention-based COmic model. Each heat-map shows the attention weights for each of the 50 pathways for three different patients. The model was trained on the GSE11121 cohort. Patient 1 was correctly classified to have a metastasis free survival (DMFS) above five years. Patient 2 was correctly classified to have a DMFS below five years. Patient 3 was wrongly classified to have a DMFS above five years while the DMFS of patient 3 was actually below five years. More examples can be found in the supplement. \textbf{D:} Mean training time of pooling-based and attention-based COmic models for differently sized datasets. The number of samples is 100, 1000, 10000, and 100000, respectively. Training was repeated five times per dataset and the stars represent the mean training time. The blue and green lines show the results for a fixed batch size of 32 samples per batch. The red and yellow lines show the results for an adaptive batch size of 1\% of the dataset size (i.e., the batch size was 1 for the dataset with 100 samples and 1000 for the dataset with 100000 samples). Each model was trained for 200 epochs.}
    \label{fig:res_singleOmics}
\end{figure}

\begin{table}
    \centering
    \caption{Total number of patients, number of patients in each class, and sources of the datasets used in single- and multi-omics prediction experiments. The first six rows contain information about the single-omics datasets used to train COmic models and compare the results to previously published methods. The last line contains information about the METABRIC multi-omics dataset used in multi-omics prediction experiments. The single-omics classes are DMFS/RFS below 5 years and above 5 years while the multi-omics classes are RFS NO and RFS YES.}
    \begin{tabular}{lcccl}
        \toprule
        Dataset & Patients & \multicolumn{2}{c}{DMFS/RFS} & Source \\
         &  & $<$ 5y / NO & $\ge$ 5y / YES &  \\
        \midrule
        GSE11121 & 181 & 28 & 153 & {\citep{schmidt2008humoral}} \\
        GSE1456 & 153 & 34 & 119 & {\citep{pawitan2005gene}} \\
        GSE2034 & 275 & 93 & 182 & {\citep{wang2005gene}} \\
        GSE2990 & 158 & 42 & 116 & {\citep{sotiriou2006gene}} \\
        GSE4922 & 228 & 69 & 159 & {\citep{ivshina2006genetic}} \\
        GSE7390 & 191 & 56 & 135 & {\citep{desmedt2007strong}} \\
        \\
        METABRIC & 1980 & 803 & 1177 & {\citep{curtis2012genomic}} \\
        \bottomrule
    \end{tabular}
    \label{tab:sets}
\end{table}


\section{Experiments on Cancer Benchmark Data}

To assess the performance and interpretation capabilities of our COmic models we use publicly available cancer benchmarks. The evaluation involves tasks on single-omics data as well as multi-omics data.

\subsection{Single-Omics Prediction on Breast Cancer Benchmark Cohorts}
\label{sec:single}

We trained COmic models on six different public breast cancer Affymetrix HGU133A microarray datasets (GSE11121, GSE1456, GSE2034, GSE2990, GSE4922, and GSE7390) that were previously used to benchmark knowledge-based classification methods that use interaction network priors. The task was to predict for each patient if metastasis free survival (DMFS) or relapse free survival (RFS) exceeded five years. On GSE11121 and GSE4922, the end point was DMFS while RFS was considered for all other cohorts. Details about the datasets can be found in Table \ref{tab:sets}. Both, pooling-based and attention-based COmic models, used 50 different pathways to build kernel layers with 30 anchor points each. We used the Laplacians derived from a merge between KEGG pathways and Pathway Commons that were publicly released by Manica and colleagues (\cite{manica2019pimkl}, see original manuscript and corresponding supplementary material for details). Furthermore, we used gated attention together with an attention dimension of 128 for the attention-based COmic models. Networks were trained for 200 epochs with the Adam optimizer {\citep{kingma2014adam}} using the class-balanced loss function {\citep{cui2019class}}. The batch size was set to 32. All models presented in this work were trained on a single NVIDIA GeForce GTX 1080 Ti. We used the area under the receiver operating characteristic (auROC) as our performance measure to be comparable to previously published results on the benchmarks. Competitors' performances shown in Figure \ref{fig:res_singleOmics}A are taken from {\citep{manica2019pimkl}}, for the PIMKL model, and {\citep{cun2012prognostic}}, for all other competitors. 

As shown in Figure \ref{fig:res_singleOmics}A, COmic models either outperformed competitors or performed similar to previously published methods. Notably, the globally interpretable pooling-based COmic models were able to achieve a small improvement in terms of auROC compared to all other models. On the other hand, the locally interpretable attention-based models achieved a similar performance as the previously best-performing model, PIMKL. We derived exemplary visualizations to evaluate the interpretation capabilities of COmic models. Since the pooling-based variant learns a molecular signature by weighting each pathway, we assessed the stability of this signature across the six breast cancer benchmarks. Each box in Figure \ref{fig:res_singleOmics}B represents one of the 50 pathways used for the prediction task and are created from the six corresponding weights learned by the models trained on the different datasets. The pathway signature remains quite stable over the six different datasets and high (absolute) weights are associated with known cancer pathways like androgen response {\citep{pietri2016androgen}}, hedgehog signaling {\citep{jamieson2020hedgehog}}, notch signaling {\citep{farnie2007mammary}}, and MYC target {\citep{xu2010myc}}. With the introduction of attention-based COmic models, we introduce models with the capability of providing local interpretations, i.e., visualizations that provide insights into the decision process for a specific sample. We show an exemplary visualization of attention weights for three different, randomly chosen patients in the GSE11121 dataset in Figure \ref{fig:res_singleOmics}C. For patient 1, the DMFS was correctly predicted to exceed 5 years. Patient 2 was correctly classified to have a DMFS below 5 years and patient 3 was wrongly classified to have a DMFS above 5 years while the actual DMFS of patient 3 was shorter than 5 years. The highest attention weights are associated with known cancer pathways. For patient 2, the highest amount of attention is given to hedgehog signaling. Androgen response gets the highest attention for patient 1 and 3. More examples can be found in the supplement.

One key advantage of artificial neural networks over kernel methods is their applicability on datasets with a vast number of samples. In the following, we will investigate, if our kernel networks provide the same applicability to large-scale datasets. Thus, we created simulated omics datasets of four different sizes: 100 samples, 1000 samples, 10000 samples, and 100000 samples. We then repeatedly trained pooling-based and attention-based COmic models on each simulated dataset five times and calculated the mean training time. Figure \ref{fig:res_singleOmics}D shows the results. Since the batch size is usually chosen based on the number of samples in the training set, we calculated the mean training time for two different batch sizes. The blue and green lines show the training times of models trained with a fixed batch size of 32 samples per batch. The red and yellow lines show the training times of models with an adaptive batch size of one percent of the total sample count, i.e., each batch included a single sample, in case of the smallest simulated dataset, and 1000 samples , in case of the largest simulated dataset. The results show that COmic models can be easily trained on datasets with several hundreds of thousands of samples with the training time being linear depended on the number of samples. Furthermore, choosing an appropriate batch size can improve the training time by more than 50\% on large-scale datasets.

\subsection{Multi-Omics Prediction on the METABRIC Benchmark Cohort}
\label{sec:multiRes}

\begin{figure}
    \centering
    \includegraphics[width=0.7\textwidth]{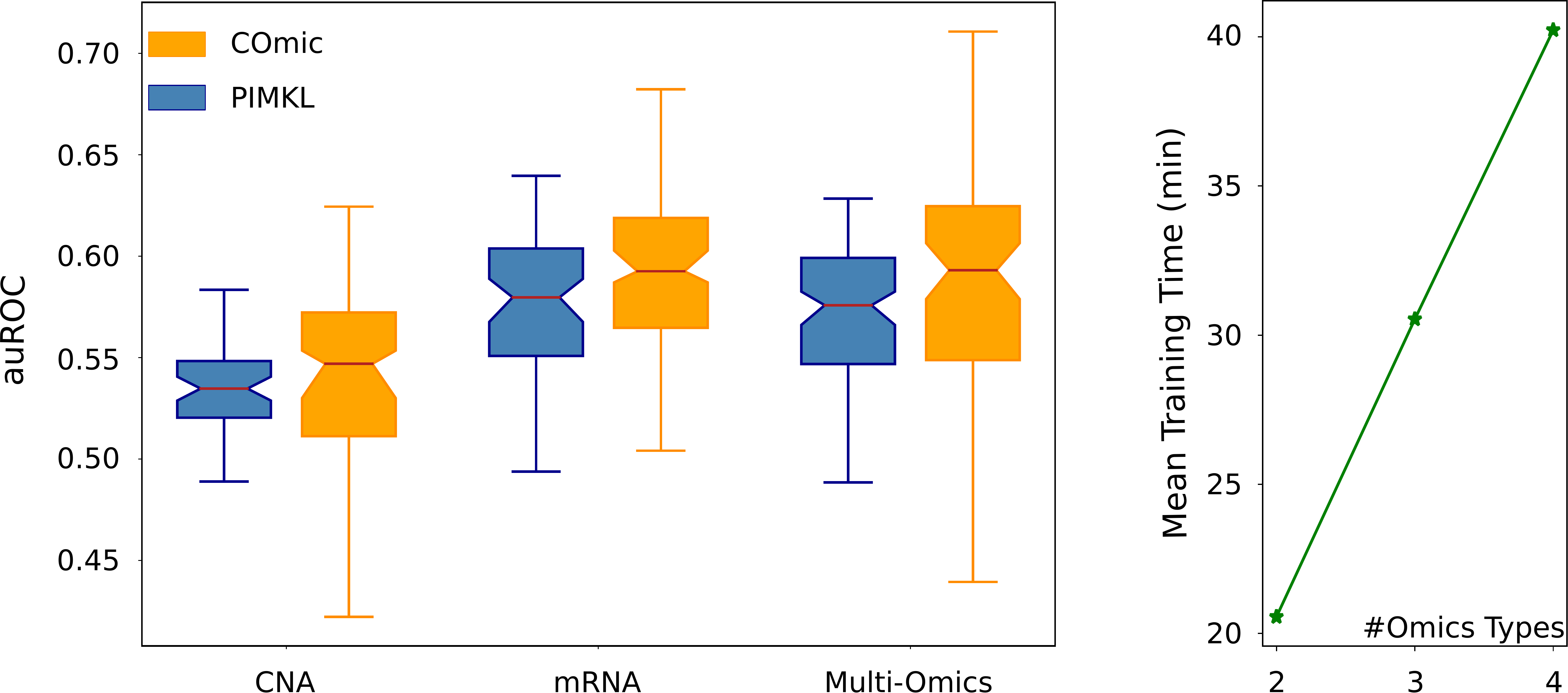}
    \caption{\textbf{Left}: Cross-validation performance of COmic and PIMKL models on the METABRIC cohort. The boxplots show the ten mean auROC validation scores of a 10-times repeated 10-fold cross-validation. Performances of COmic models are shown in orange. The center line of each box indicates the median. The height of the boxes represents the inter-quartile range (IQR) with the upper and lower whiskers set to 1.5 times the IQR. Notches represent the confidence interval (CI) around the median and were calculated using bootstrapping with 10000 iterations. \textbf{Right}: Mean training time of multi-omics COmic models. Artificial datasets with 2, 3, and 4 different omics modalities and 1000 data points per modality were investigated. Batch size and number of epochs were fixed as described for the METABRIC cohort. Training of models was repeated five times and mean times are indicated by stars.}
    \label{fig:res_multiOmics}
\end{figure}

Since our proposed kernel layer can be incorporated into any ANN, COmic models can be flexibly expanded to multi-omics datasets. One possibility is to directly add the pathway kernels for the additional omics datatypes to the kernel layer, thereby increasing the number of graph Laplacians in the kernel layer. Another simple approach is to create sub-networks for each omics type, i.e., combine the output of pooling- or attention-based single-omics COmic models with a simple fully connected network. There are numerous other ways to expand COmic models to multi-omics data and, since our proposed approach is knowledge-driven, the individual solution has to be selected with the context of the data in mind. Similar to the authors of PIMKL, we chose the METABRIC cohort to investigate the practicality of applying COmic models to multi-omics datasets. The METABRIC breast cancer cohort contains gene expression (mRNA) and copy number alteration (CNA) data. We performed the same prediction task as in {\citep{manica2019pimkl}}, i.e., using molecular measures to predict whether a patient had recurrent cancer.

While we used pooling-based models with the same hyperparameters as described above for the single-omics prediction experiments, the creation of a multi-omics COmic model for the METABRIC cohort had to be done with caution. The issue with CNA data is that this datatype is tremendously sparse. To compensate for this sparseness, we build a network that used pooling-based kernel layers to compute an embedding for each datatype that is robust enough for sparse data and afterwards used a gated attention layer as described in section \ref{sec:interpret} to make the prediction. The pooling-based kernel layers used the same hyperparameters as the pooling-based COmic models in the single-omics experiments. For the gated attention layer we used an attention dimensionality of 4. Networks were trained for 200 epochs with the Adam optimizer {\citep{kingma2014adam}} using the class-balanced loss function {\citep{cui2019class}}. The batch size was set to 32. We used the auROC as our performance measure to enable the comparison to previously published PIMKL results on the METABRIC cohort. Additionally, we investigated the computational efficiency of COmic with regard to an increasing number of omics modalities.

The results of our experiments on the METABRIC cohort can be found on the left side of Figure \ref{fig:res_multiOmics}. The shown PIMKL performance is taken from {\citep{manica2019pimkl}}. As expected, neither PIMKL nor COmic achieved good performance on the sparse CNA data. COmic models slightly outperformed PIMKL on the single-omics prediction task using gene expression data. While PIMKL shows a slightly decreased performance on the multi-omics prediction task, COmic seems to have the same performance on the multi-omics data as on the gene expression data alone with an increase in variance. The runtime analysis (right side of Figure \ref{fig:res_multiOmics}) shows a linear dependency on the number of omics modalities.

\section{Discussion}
Kernel methods allow to induce prior knowledge into a prediction task resulting in increased robustness and the introduction of interpretation capabilities. In this work we propose COmic, a method to incorporate pathway-induced kernel functions into convolutional kernel networks. We are able to create learning models that can be robustly trained on small-scale datasets and scale very well with the number of samples. Thus, they can be efficiently applied to large datasets with hundreds of thousands of samples. Furthermore, our models provide global and local interpretations of predictions made on molecular measures due to the pathway-induced kernel function.

We used six different breast cancer cohorts to compare the performance of COmic models to previously proposed methods that use prior knowledge for prediction tasks with molecular measures as input data. The results presented in Figure \ref{fig:res_singleOmics}A show that our method reaches state-of-the-art performance on classifying patients based on their DMFS/RFS from gene expression data: Compared to the considered competitors, COmic performs similar or even better. However, COmic models have the advantage that the time needed to train a model scales linearly with the number of samples (see Figure \ref{fig:res_singleOmics}D). This enables the use of COmic models on datasets with hundreds of thousands of samples. We provide evidence that our method can be readily applied on large datasets by training models on simulated single-omics data with sizes ranging from 100 to 100,000 samples. Although datasets and patient cohorts used in computational biology and medicine traditionally have smaller sample sizes, high-throughput methods and the nowadays more frequently used big data paradigm will result in increasing sample counts in biological and medical datasets. At this day, TCGA already contains data from more than 85,000 patients. While methods that can deal with large datasets are usually deployed as black-box models, our method provides increased insight into the decision making process.

Using single-omics datasets strongly limits the decision process for diagnosis of a majority of diseases. Nowadays, it is well known that multi-omics information has to be incorporated to get a complete image of the pathomechanism causing a certain disease. Therefore, methods that are limited to a single datatype face serious constraints if employed as a decision-support system or to deepen knowledge about a pathomechanism. Our proposed method does not face the limitation of only using single-omics data as we show in our experiment with the METABRIC multi-omics cohort. The results show again that our method improves single-omics prediction as demonstrated by the performance on the gene expression data. The lower performance that both methods, PIMKL and COmic, show on the copy number alteration data can be explained by the sparseness of CNA data. Sparse data poses serious problems for prediction models {\citep{li2016convergence}} and both methods are not specifically designed for sparse data. However, we can show in Figure \ref{fig:res_multiOmics} that COmic models are able to achieve slightly better performance than PIMKL models on the multi-omics prediction task. This indicates that our approach could be advantageously used on multi-omics data, while the flexibility of the architecture (as described in section \ref{sec:multiRes}) enables researchers to tailor COmic models for specific datasets using domain expertise.

Computing an interpretation of a machine learning method, either with \textit{post-hoc} methods or through intrinsically interpretable models, is beneficial if and only if the interpretation serves a purpose. This purpose cannot be defined in general as it is highly dependent on the task, the data, and the user that is presented with the obtained interpretation. For the presented experiments, we investigated if the inherent interpretation capabilities of our COmic models are able to learn biological meaningful concepts directly from data. First, we considered the global interpretation capabilities of COmic. Here, COmic models assign a weight to each of the used pathways and the weights reflect the role that each pathway plays in classifying an input sample, i.e., a patient. We trained COmic models on six different single-omics breast cancer cohorts. Since we expect the biological processes in the cohorts to share high similarities, the weight signatures of all models should be similar if the COmic method is able to learn meaningful pathways from data. As shown in Figure \ref{fig:res_singleOmics}B, this assumption is indeed well fulfilled with all six models having similar weight signatures. Furthermore, pathways with a high weight assigned to them are mainly known cancer-related pathways. Therefore, COmic models are able to learn biological meaningful pathway weights. Although the previously published PIMKL method also has a global interpretation capability, our method is able to learn pathways that are important for both, the negative and the positive class, due to the fact that the learned weights can be positive or negative. PIMKL only learns positive weights. 

While global interpretation is useful to gain insights into a dataset, local interpretation can be used to get insights into the decision a model makes for a specific input. Attention-based COmic models can provide this insight utilizing the attention weights that are computed for each input sample separately. These weights directly determine the influence that each pathway has on the decision made by the model. We can visualize these influence using a heatmap (as shown in Figure \ref{fig:res_singleOmics}C) to quickly see which pathways played an important role in the decision made. We randomly selected three samples from the GSE11121 dataset to evaluate if the attention weights are biological meaningful. Similar to the weight signatures of the globally interpretable COmic models, the attention weights of the locally interpretable COmic models highlighted known cancer-related pathways. Interestingly, the selected patient with a DMFS below five years has attention weights that are strongly focused on a single pathway. This is true for all correctly classified patients with a DFMS below five years (see supplement). On the other hand, patients with a DMFS below five years that were wrongly classified to have a DMFS above five years show attention weight patterns similar to those of patients with a DMFS above five years (see patient 3 in Figure \ref{fig:res_singleOmics} and additional examples in the supplement). This could indicate that the wrongly classified patients exhibit a different mechanism causing a DMFS below five years, compared to the correctly classified ones, which was not learned by the model. The local interpretation capabilities of COmic models can help to directly show possible directions to further investigate the data. Furthermore, the results of our experiments strongly suggest that both COmic model types are able to generate biologically meaningful interpretations. We chose heatmaps to visualize attention weights, since it appeared convenient for the considered prediction task on the studied dataset. However, different forms of explanations can be computed with attention weights, e.g., counterfactual explanation {\citep{tran2021counterfactual}} and adversarial explanation {\citep{kitada2021attention}}. The most suitable form of explanation is highly dependent on the application, target user group, and the goal aimed at by the explanation. Therefore, the chosen visualization should be understood as an example and not a general application recommendation.

COmic models have a few hyperparameters that can be optimized using appropriate methods like, e.g., grid search or random search. These hyperparameters include the number of anchor points, the attention type, the dimensionality of the attention layer's parameters $V$ and $U$, and the choice of pathways used for kernel layers. Furthermore, different initialization procedures for the anchor points can be explored, e.g., a parameter-free clustering that combines initializing anchor points with optimizing the number of anchor points for each pathway-induced kernel layer. We recommend to explore hyperparameter optimization when applying COmic models. However, minimizing energy consumption is a pressing concern that should be considered in every line of research nowadays. Therefore, we limited the computations performed for this work to the minimum required to support our claims. The hyperparameters for all models presented in this work were chosen by combining prior experience about kernel networks with domain expertise. Interestingly, this computation-free approach to hyperparameter selection already leads to competitive performance of our method on the considered prediction tasks.

\section{Conclusion}
The introduced convolutional omics kernel networks utilize prior knowledge by pathway-induced kernel functions to provide robust end-to-end learning on small- to large-scale molecular measure datasets. Furthermore, utilizing pathway-induced kernel functions makes our method intrinsically interpretable with the ability to provide global and local interpretations.

We show the competitive performance of our method on six different single-omics breast cancer cohorts while providing new interpretation capabilities that exceed the possibilities of previously proposed methods. Furthermore, we show that COmic models can be readily adapted to multi-omics datasets.

On a larger scale, we show that incorporating a carefully crafted kernel function into an artificial neural network allows to robustly train ANNs on small-scale datasets as they frequently occur in computational biology and medicine. On the other hand, our method enables scientist to utilize kernel functions for large datasets as they arise more frequently with the increasing use of high-throughput methods and big data.

\begin{ack}
Funded by the Deutsche Forschungsgemeinschaft (DFG, German Research Foundation) under Germany’s Excellence Strategy – EXC number 2064/1 – Project number 390727645. This research was supported by the German Federal Ministry of Education and Research (BMBF) project ’Training Center Machine Learning, Tübingen’ with grant number 01|S17054. This work was supported by the German Federal Ministry of Education and Research (BMBF): Tübingen AI Center, FKZ: 01IS18039A.
\end{ack}

\bibliographystyle{apalike}
\bibliography{references}

\newpage

\appendix

\section{Societal and Environmental Impact}
Medical data are notoriously biased against minorities and there are numerous examples of machine learning models that learn this biases and have a severe deterioration in performance with regard to minorities (see e.g., \cite{morley2020ethics}). We did not include a statement on the societal impact of our work into the main manuscript due to the fact that we did not have meta information about ethnicity of patients included in the used benchmarks. Therefore, it was not feasible to investigate if the prediction performance and interpretation capabilities of COmic models change for minorities. However, we encourage researchers that want to apply COmic models on real-world data to investigate potential bias in their results.

All experiments were conducted using a single NVIDIA GeForce GTX 1080 Ti GPU. All experiments together required a approximated total of 146 hours of computing time. This resulted in total emissions of 15.77 kg CO$_2$e, which is equivalent to burning 7.9 kg of coal. To compensate this emissions, 0.26 tree seedlings have to sequester carbon for 10 years. These estimations were calculated using the Machine Learning Impact calculator\footnote{https://mlco2.github.io/impact/} by Lacoste and colleagues \citep{lacoste2019quantifying}.

\section{Additional Examples of the Local Interpretation Abilities of Attention-Based COmic Models}

\begin{figure}
    \centering
    \includegraphics[width=\textwidth]{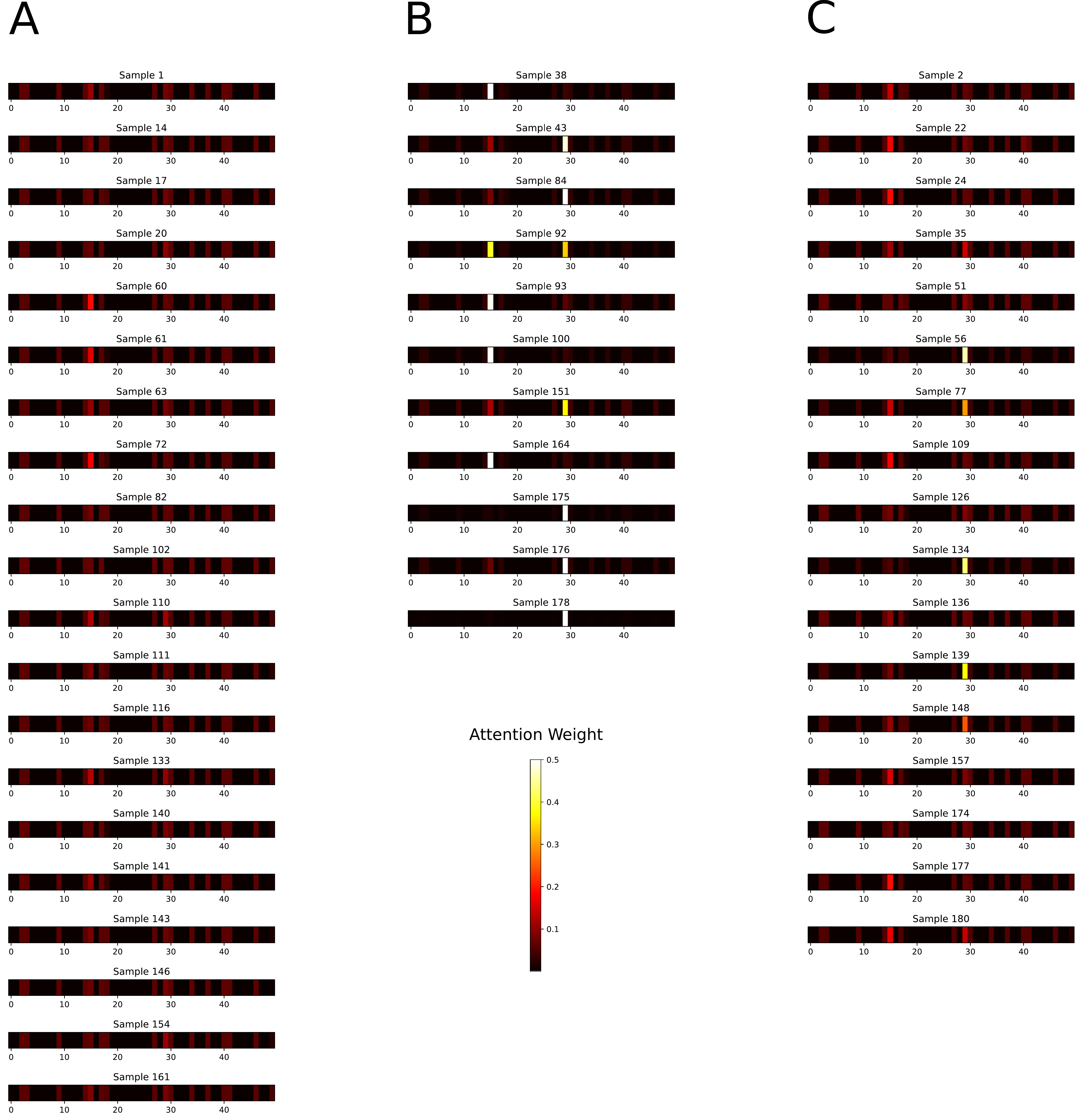}
    \caption{Additional examples of the local interpretation capability of attention-based COmic models. \textbf{A}: Randomly selected samples from the GSE11121 cohort that belong to patients that had a metastasis free survival of more than five years. All samples represent patients that were correctly predicted to have a metastasis free survavival of more than five years. \textbf{B}: All samples from the GSE11121 cohort that belong to patients that had a metastasis free survival of less than five years and were correctly predicted to have a metastasis free survival of less than five years. \textbf{C}: All samples from the GSE11121 cohort that belong to patients that had a metastasis free survival of less than five years and were wrongly predicted to have a metastasis free survival of more than five years.}
    \label{fig:additional_local}
\end{figure}

We present additional examples of the local interpretation capabilities of our attention-based COmic models. Figure \ref{fig:additional_local} \textbf{A} shows a set of 20 randomly selected patients from the GSE11121 cohort that were correctly predicted to have a metastasis free survival of more than five years. Figure \ref{fig:additional_local} \textbf{B} shows all patients from the GSE11121 cohort that were correctly predicted to have a metastasis free survival of less then five years. Figure \ref{fig:additional_local} \textbf{C} shows all patients from the GSE11121 cohort that were wrongly predicted to have a metastasis free survival of more than five years when their real metastasis free survival was less than five years. The order of the pathway attention weights are the same as in the corresponding figure in the main manuscript.

We can see that the group of correctly classified patients with a metastasis free survival of more than five years show a similar pattern with a relative high attention weight on many different pathways. On the other hand, patients that were correctly classified to have a metastasis free survival of less than five years show attention weights that are focused on a single pathways or two pathways at most. Furthermore, the focused pathways are always the same, hedgehog signalling (pathway 29) and androgen response (pathway 16). The hedgehog signalling pathway is known to be important in tumor metastasis \citep{li2006snail}. Furthermore, the androgen receptor plays an important role in the breast development cycle and also is known to affect breast cancer development and metastasis \citep{pietri2016androgen,arce2014complete}. The focus of attention weights on these two pathways in patients with shorter metastasis free survival reflects the known facts about the role that both pathways play in this process. However, there is an interesting pattern that can be observed for patients with a metastasis free survival of less than five years that were wrongly classified to have a metastasis free survival above five years. While some have the clear focus on either hedgehog signalling or androgen response and, therefore, strongly suggest that they are simple wrongly classified patients, most of them shows similar attention weight patterns to patients with an actual metastasis free survival above five years. This could hint at the fact that the group of wrongly classified patients with strong attention weight similarities to patients with longer metastasis free survival have a different mechanism that causes metastasis than correctly classified patients with short metastasis free survival. And this different mechanism is not learned by the model. In any case, the interpretation indicates that there are different groups of wrongly classified patients and these groups should be investigated further. These findings support our claim of the benefits of using intrinsically interpretable models due to the fact that these different groups in wrongly classified patients can be easily noticed directly from the model without further polluting the prediction task with necessary assumptions and additional computations for \textit{post-hoc} interpretation models.

\end{document}